\documentclass{aa}

\usepackage{graphics}

\begin{document}

\title{Optical polarimetric monitoring of the type II-plateau SN~2005af~\thanks{Based on observations obtained at the {\it Observat\'orio do Pico dos Dias}, LNA/MCT, Itajub\' a, Brazil.}}
\author{A. Pereyra\inst{1}
\and A. M. Magalh\~aes\inst{1}
\and C. V. Rodrigues\inst{2}
\and C. R. Silva\inst{3}
\and R. Campos\inst{3}
\and G. Hickel\inst{4}
\and D. Cieslinski\inst{2}
}

\offprints{A. Pereyra, \email{antonio@astro.iag.usp.br}}

\institute{
Departamento de Astronomia, IAG, Universidade de S\~ao Paulo, Rua do Mat\~ao
 1226, S\~ao Paulo, SP, 05508-900, Brazil
\and
Instituto Nacional de Pesquisas Espaciais/MCT, Avenida dos Astronautas 1758,
 S\~ao Jos\'e dos Campos, SP, 12227-010, Brazil
\and
Laborat\'orio Nacional de Astrof\'{\i}sica/MCT, CP 21, Itajub\'a, MG, 37500-000,
 Brazil
\and
IPD - UNIVAP, Av. Shishima Hifumi, 2911, Urbanova, S\~ao Jos\'e dos Campos, SP,
 12244-000, Brazil
}

\date{Received dd-mm-yy / Accepted dd-mm-yy}

\abstract
{}
{Core-collapse supernovae may show significant polarization that implies non-spherically symmetric explosions. We observed the type II-plateau SN~2005af using optical polarimetry in order to verify whether any asphericity is present in the supernova temporal evolution.}
{We used the IAGPOL imaging polarimeter to obtain optical linear polarization measurements in ${\it R}$ (five epochs) and ${\it V}$ (one epoch) broadbands. Interstellar polarization was estimated from the field stars in the CCD frames. The optical polarimetric monitoring began around one month after the explosion and lasted $\sim$30 days, between the plateau and the early nebular phase.}
{The weighted mean observed polarization in ${\it R}$ band was [1.89~$\pm$~0.03]$\%$ at position angle (PA) 54\fdg1. After foreground subtraction, the level of the average intrinsic polarization for SN~2005af was $\sim$0.5$\%$ with a slight enhancement during the plateau phase and a decline at early nebular phase. A rotation in PA on a time scale of days was also observed. The polarimetric evolution of SN~2005af in the observed epochs is consistent with an overall asphericity of $\sim$20$\%$ and an inclination of $\sim$30$\degr$. Evidence for a more complex, evolving asphericity, possibly involving clumps in the SN~2005af envelope, is found.}
{}


\titlerunning{Optical polarimetric monitoring of SN~2005af}
\authorrunning{Pereyra et al.}
\maketitle

\section{Introduction\label{intro}}

SN~2005af was discovered by Jacques \& Pimentel (\cite{ja05}) on Feb.~8.22 (UT) and is located 407$\arcsec$ West and 351$\arcsec$ South of the center of NGC~4945. Filippenko \& Foley (\cite{fi05}) reported spectroscopy on Feb.~12, around one month after the explosion. They showed that SN~2005af is a type II-plateau with well-developed P-Cyg profiles of H, Fe II, Ca II and other species.

In the past few years broadband polarimetry and spectropolarimetry have been shown to be unique tools to study the geometry of supernovae (SNe) explosions. In particular, core-collapse SNe show significant levels of polarization, implying that asphericity is present in these events (Wang et al. \cite{wa96}, Leonard et al. \cite{le01}, Leonard \& Filippenko \cite{lefi01}, Wang et al. \cite{wa02}). The polarization level in core-collapse SNe may be associated with the mass of the hydrogen envelope that remains intact at the time of explosion (for a review see Leonard \& Filippenko, \cite{le05}). In this context, type II-plateau SNe (SNe II-P), with progenitors with large and intact envelopes, tend to have very low polarization. However, when part (SN IIb, SN IIn) or all (SN Ib) of the hydrogen was lost before the explosion, important polarization levels are expected. Up to now, the intrinsic polarization for SNe II-P events was found to be lower than  0.5\%$-$0.6\% and evidence of a polarization enhancement between the plateau and the early nebular phase exists  (Leonard et al. \cite{le01,le06}). Temporal evolution coverage in SNe II-P events is hence crucial to have a better understanding about the asphericity variation when the deepest layers of the ejecta are revealed.

In this letter we present an optical polarimetric monitoring of the type II-plateau SN~2005af between Feb.~14 and Mar.~12, starting approximately one month after the explosion. The observations and data reduction are presented in Sect. \ref{data}. The estimates of interstellar and intrinsic polarizations are derived in Sects. \ref{foregroundP} and  \ref{intrinsicP}, respectively. Discussion and final comments are presented in Sect. \ref{comm}.

\section{Observations\label{data}}

\begin{table*}
\caption{Log of observations.}
\begin{tabular}{lccccccccc}
\hline \hline
Date   & Telescope & Filter & Waveplate & IT &   Pos. & {\it P}$_{\rm obs}$ & $ \theta$$_{\rm obs}$  & {\it P}$_{\rm int}$ & $ \theta$$_{\rm int}$\\
(2005, UT middle) &  &   & used & (s) &  & (\%) & (degrees) & (\%) & (degrees) \\
\hline
Feb. 14.230 & 60 cm & {\it R} & $ \lambda $/2 & 60 &  8  & 1.80 (0.07) &  53.4 (1.2) & 0.39 (0.08) & 26.2 (5.9)  \\
Feb. 17.231 & 60 cm & {\it R} & $ \lambda $/2 & 80 & 16  & 1.86 (0.06) &  51.9 (0.9) & 0.50 (0.07) & 25.4 (4.0) \\
Feb. 22.240 & 60 cm & {\it R} & $ \lambda $/2 & 80 & 8   & 1.98 (0.13) &  52.1 (1.8) & 0.57 (0.13) & 30.6 (6.6)  \\
Feb. 24.316 & 60 cm & {\it R} & $ \lambda $/2 & 160 & 8  & 2.00 (0.08) &  54.1 (1.2) & 0.51 (0.09) & 37.3 (5.0) \\
Mar. 12.215 & 160 cm & {\it R} & $ \lambda $/4 & 20 & 16 & 1.92 (0.07) &  58.1 (1.0) & 0.32 (0.08) &  52.9 (6.8) \\
Mar. 12.231 & 160 cm & {\it V} & $ \lambda $/4 & 40 & 16 & 2.42 (0.20) &  55.4 (2.4) & 0.82 (0.20) &  47.8 (7.2) \\
\hline
\end{tabular}
\\
The errors are in parenthesis.
\label{log}
\end{table*}

\begin{figure}
\resizebox{\hsize}{!}{\includegraphics{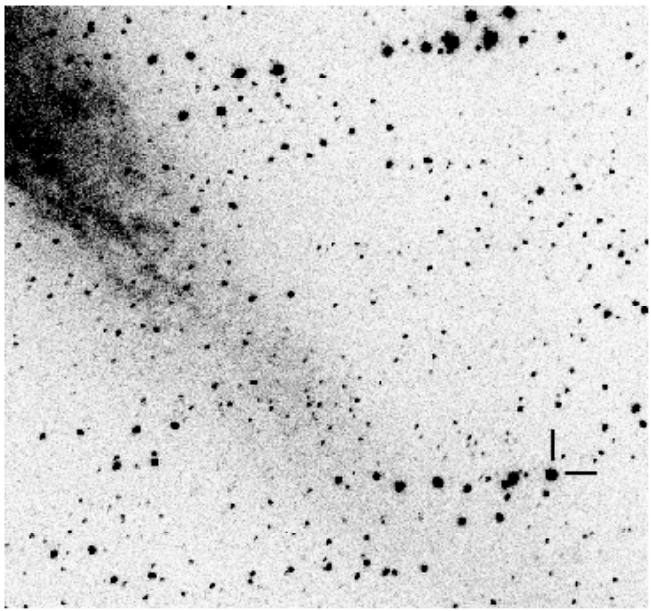}}
\caption{Image of the SN~2005af field obtained with the IAGPOL polarimeter in  {\it R} band on Feb. 14.230. Each object has two images, each orthogonally polarized to the other, due to the polarimeter's calcite prism. The south-west
 part of NGC~4945 is shown at the left of this image. The supernova, located 407$\arcsec$ West and 351$\arcsec$ South of the center of NGC~4945, is indicated. North is to the top and East is to the left.}
\label{sn}
\end{figure}

The observations were made using IAGPOL, the IAG imaging polarimeter (Magalh\~aes et al. \cite{ma96}), at the Observat\'orio do Pico dos Dias (OPD), Brazil. Additional details about this polarimeter and the experimental procedure can be found in Pereyra (\cite{pe00}) and Pereyra \& Magalh\~aes (\cite{pe02,pe04,pe05}).

Table~\ref{log} shows the log of observations. The f/13.5 Cassegrain focus of the 0.6m IAGUSP telescope was used between Feb. 14 and Feb. 24 (see Fig.~\ref{sn}). The last observations on Mar. 12 were made at the f/10 Cassegrain focus of the 1.6m telescope. 

We used IAGPOL in linear polarization mode with an achromatic $\lambda$/2-waveplate in all the runs except the last one (Mar. 12) when a superachromatic  $\lambda$/4-waveplate was employed. The {\it R} filter was used in all epochs and an additional {\it V } filter measurement was gathered on Mar. 12. The integration time {\it per} waveplate position (IT) and the number of positions used in each measurement (Pos.) are indicated in Table~\ref{log}. The observations were made using a 1024$\times$1024 CCD, covering $\sim$10$\arcmin\times$10$\arcmin$ and 5$\arcmin\times$5$\arcmin$ at the 0.6m and 1.6m telescopes, respectively.

For the data reduction process we follow the procedure indicated in Pereyra \& Magalh\~aes (\cite{pe02}) using the PCCDPACK package (Pereyra \cite{pe00}), written for the IRAF\footnote{IRAF is distributed by the National Optical Astronomy Observatory, which is operated by the Association of Universities for Research in Astronomy, Inc., under cooperative agreement with the National Science Foundation.} environment. We used observations of polarized standard stars taken each night to convert the resulting instrumental polarization position angle (PA) to the equatorial system. The instrumental polarization obtained from observations of unpolarized stars was smaller than 0.16$\%$, and no such correction was applied.

\begin{figure}
\resizebox{\hsize}{!}{\includegraphics{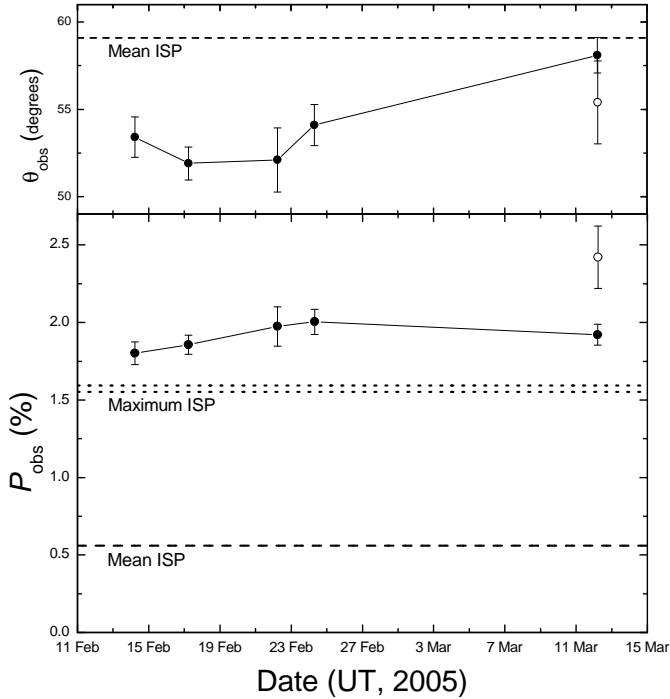}}
\caption{Temporal evolution coverage for the observed polarization of SN~2005af between Feb. 14 and Mar. 12. The bottom panel is the polarization value and the top panel is the position angle. {\it R} and {\it V} band data are in black
 and open dots, respectively. The dashed lines indicate the mean ISP calculated using the field stars (0.56\% at PA 59$\fdg$1). The dotted lines indicate the maximum ISP in {\it R} (1.61\%, bottom) and {\it V} (1.65\%, top) bands toward the position of SN2005af assuming {\it P}$_{V,max}$ = 9{\it E}({\it B}$-${\it V}) and {\it P}$_{R,max}$
 from extrapolation of the Serkowski's law (with $ \lambda $$_{max}$ = 0.55 $ \mu$m).}
\label{evol}
\end{figure}

The observed polarization ({\it P}$_{\rm obs}$ and $ \theta $$_{\rm obs}$) for SN~2005af in five epochs in the {\it R} band (Feb. 14, 17, 22, 24 and Mar. 12) and one epoch in the {\it V} band (Mar. 12) is shown in Table~\ref{log}. Light curves for SN~2005af on {\it u}\arcmin{\it g}\arcmin{\it r}\arcmin{\it i}\arcmin{\it B}{\it V} filters are available at CSP\footnote{Carnegie Supernova Project, http://csp1.lco.cl/$\sim$cspuser1
 /images/2004\underline{ }2005\underline{ }opt\underline{ }lightcurves/SN05af.html} site. The first four of our measurements (Feb. 14, 17, 22 and 24) are coincident with the end of the plateau phase and the last one (Mar. 12) with the beginning of the nebular phase. Filippenko \& Foley (\cite{fi05}) dated Feb. 12 as one month after the explosion; the plateau phase for SN~2005af then lasted $\sim$60 days.

We show in Fig.~\ref{evol} our polarimetric temporal evolution coverage for SN~2005af. The overall, weighted mean observed polarization in ${\it R}$ band was [1.89~$\pm$~0.03]$\%$ at PA 54\fdg1. The maximum variations of {\it P}$_{\rm obs}$ and $ \theta $$_{\rm obs}$ are 2.5 and 5.1 times larger than the mean errors of the individual measurements, respectively. Therefore, the observed increase in {\it P}$_{\rm obs}$ at the end of the plateau phase (i.e., the first four epochs) is marginal. This is not the case for $ \theta $$_{\rm obs}$ where evidence of rotation on a time scale of days seems to be present. This is confirmed if we applied the statistical criteria from Kesteven et al. (\cite{ke76}) to evaluate temporal polarization variations.  The probability\footnote{The observable is variable if {\it p} $<$ 0.1\%, possibly variable if 0.1\% $\leq$ {\it p} $\leq$ 1\% and non-variable if {\it p} $>$ 1\%.} {\it p}($\chi$$^{2}$) of exceeding $\chi$$^{2}$ by chance is 36.2\% (i.e., non-variable) and 0.02\%  (i.e., variable) for {\it P}$_{\rm obs}$ and $ \theta $$_{\rm obs}$, respectively. This finding suggests that time variability of the observed polarization is intrinsically related to the supernova (SN) event, regardless of the interstellar polarization (ISP) toward the SN.

\section{Interstellar polarization}
\label{foregroundP}

The ISP is an additive component in the SN observed polarization that must be estimated if we want to know the SN intrinsic polarization. We estimated the ISP for SN~2005af using field stars present in the CCD frames. The PCCDPACK package provides the polarization of each object in the CCD frames, as well as the error-weighted average Stokes parameters ({\it Q} and {\it U}) for all the objects with a given polarization signal-to-noise ratio ({\it P}/$ \sigma ${\it $_{P}$}). We take these parameters as the average ISP toward the SN.

\begin{figure}
\resizebox{\hsize}{!}{\includegraphics{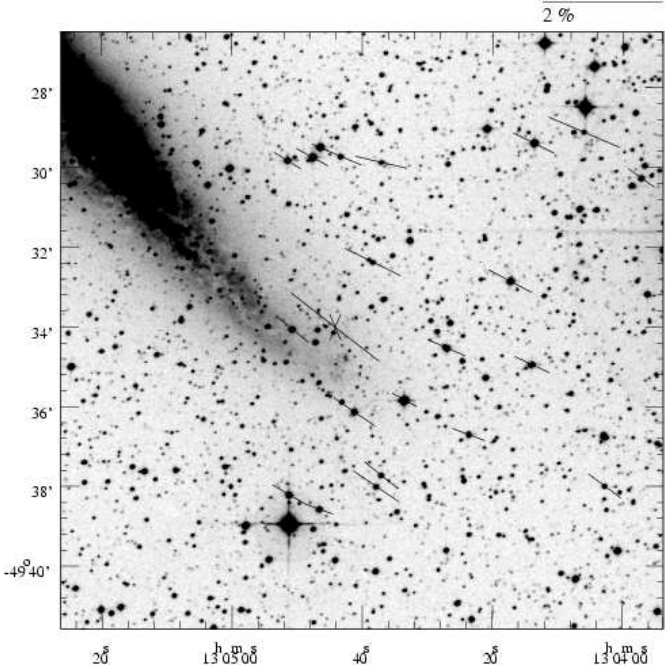}}
\caption{ISP map (in {\it R} band) for the SN~2005af field on Feb. 17. This DSS2 Red image covers 15$\arcmin\times$15$\arcmin$. Each vector represents the polarization associated with each foreground object (located at the middle of the vector). The position of SN~2005af is indicated with a X and also its polarization measured in this date is shown. The scale is shown at the top right. The coordinates are 2000.}
\label{field}
\end{figure}

Twenty-four foreground objects with {\it P}/$ \sigma ${\it $_{P}$} $>$ 5 were used to calculate the ISP and some of them were measured in more than one epoch. The weighted mean ISP from all dates except Mar. 12 was [0.56~$\pm$~0.01]$\%$ at PA 59\fdg1 (indicated by a dashed line in Fig.~\ref{evol}). As an example, in Fig.~\ref{field} we show the polarization map in the {\it R} band for the foreground objects on Feb. 17 along with the SN measurement on this date. Clearly, the SN is the object with the highest polarization in the field but its PA is very similar to the direction of the ISP. This suggests that the SN's observed polarization indeed has an ISP component. A histogram of USNO-B1 magnitudes for all the foreground objects used in our calculation indicates that the sample is complete up to 13.5 mag and probably represents close objects. If this were the case, the direction of the ISP toward the SN would be well represented by the PA estimated from the field stars but with the polarization value underestimated. 

We also checked the agglomeration of stellar polarization catalogs from Heiles (\cite{he00}) and noted that, for the twelve objects\footnote{We excluded variable, binary, emission-line and post-AGB stars in this selection.} in a 14$\degr$$\times$14$\degr$ region centered on the host galaxy (NGC~4945: {\it l}$=$305\fdg3, {\it b}$=$13\fdg3) the mean PA is (73~$\pm$~17)$\degr$. Such a value is consistent with our PA (59\fdg1) obtained from the field stars. As noted by Jacques \& Pimentel (\cite{ja05}), NGC~4945 and SN~2005af are located toward the Centaurus arm in a rather rich area of foreground Galactic stars. This is suggestive of a significant ISP component.

To verify this further, we estimated the expected maximum polarization ($\sim$9$\times${\it E}({\it B}$-${\it V}), Serkowski et al. \cite{se75}) toward the SN assuming a Galactic reddening of {\it E}({\it B}$-${\it V}) = 0.183~$\pm$~0.004\footnote{see the Dust Extinction Service available at http://irsa.ipac.caltech.edu/applications/DUST/} toward SN2005af (Schlegel et al. \cite{sc98}). The maximum polarization is then {\it P}$_{V,max}$ = [1.65~$\pm$~0.04]\%. We feel that this polarization level is more representative of the actual ISP toward SN~2005af. Therefore, the main component in the SN's observed polarization is of Galactic origin. Using Serkowski's law with $\lambda $$_{max}$ = 0.55 $ \mu $m for the ISP we estimate {\it P}$_{R,max}$ = [1.61~$\pm$~0.04]\%. These values are indicated by dotted lines in Fig.~\ref{evol} and will be used below to obtain the intrinsic polarization for SN~2005af (see the last two columns of Table~\ref{log}).

\begin{figure}
\resizebox{\hsize}{!}{\includegraphics{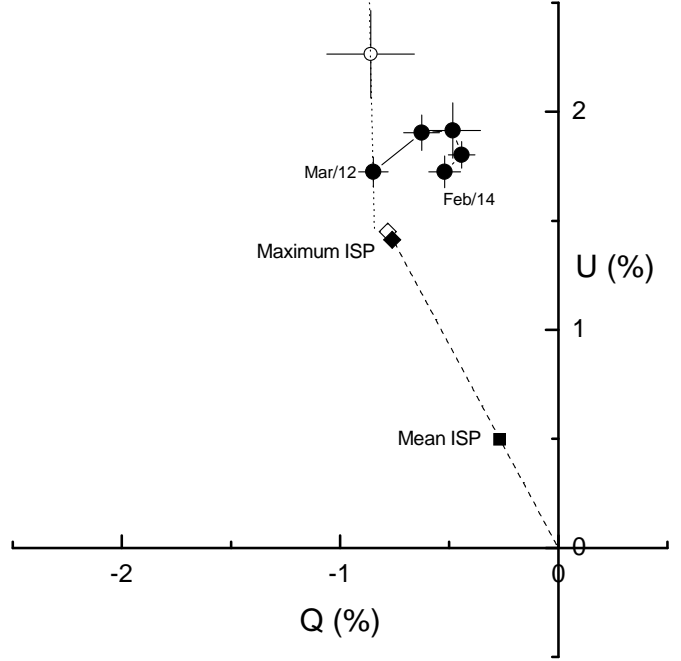}}
\caption{{\it Q}$-${\it U} diagram for SN~2005af and ISP. In black dots are the observed polarization for SN~2005af in {\it R} band data between Feb. 14 and Mar. 12. In open dots is the {\it V} band observation on Mar. 12 for SN~2005af. In black squares is the mean ISP in {\it R} band from the field stars. In black and open diamonds are the maximum ISP (in {\it R} and {\it V} bands, respectively) toward the position of SN~2005af (as in Fig.~\ref{evol}) assuming the PA is the same as that obtained using the field stars (dashed line). The dotted line indicates a linear fit to {\it R} and {\it V} SN~2005af data on Mar. 12.}
\label{qu}
\end{figure}

\begin{figure}
\resizebox{\hsize}{!}{\includegraphics{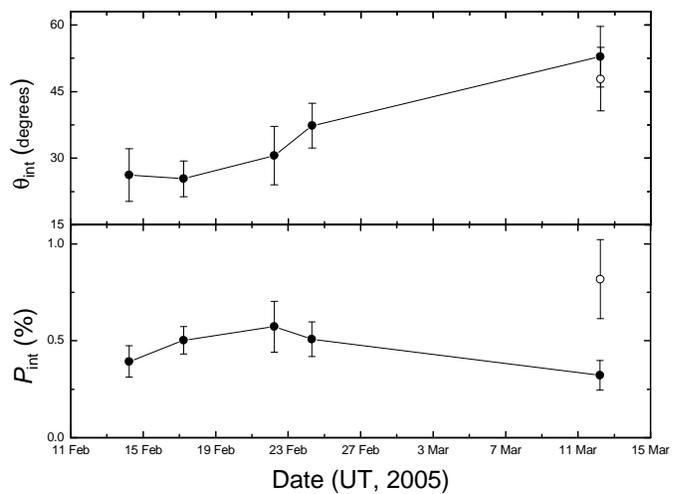}}
\caption{Temporal evolution coverage for the intrinsic  polarization of SN~2005af after foreground subtraction. {\it R} and {\it V} band data are in black and open dots, respectively. }
\label{polint}
\end{figure}

\section{Intrinsic polarization}
\label{intrinsicP}

In Fig.~\ref{qu} we plotted the {\it Q}$-${\it U} diagram of the observed temporal evolution of the SN~2005af polarization, along with the foreground estimates of Sect. \ref{foregroundP}. In such a plot, a point has polar coordinates ({\it P},~2$\theta$). The PA rotation in the {\it R} band values noted in Sect.~\ref{data} is evident in the figure. 

We can use the two color measurements on Mar. 12 to independently verify our ISP estimates. These data were taken in an interval of less than 30 minutes and must represent the instant wavelength dependence for the observed polarization at that epoch. Therefore, the observed rotation in $ \theta $$_{\rm obs}$ is not expected to impair the comparison of the PA values in each band in the Mar. 12 data. The dotted line in Fig.~\ref{qu} connects the {\it R} and {\it V} data on Mar. 12. Its intersection with the direction of ISP (dashed line) obtained from field stars is approximately coincident with our estimates of maximum polarization along the line of sight toward SN~2005af.  This finding seems to confirm that the observed polarization on Mar. 12 can indeed be decomposed into two components, foreground and intrinsic, with different position angles, namely, $\sim$59$\degr$ for the ISP and $\sim$48$\degr$ for the intrinsic component. Therefore, the interstellar polarization is well represented by the maximum polarization estimated in Sect. \ref{foregroundP}. On the other hand, the intrinsic polarization shows an approximately constant PA between the two colors on Mar. 12. This fact is consistent with the assumption that electron scattering is the main source of the intrinsic polarization in SNe (see Sect. \ref{comm}) with a flat wavelength dependence on PA. 
 
After the subtraction of the maximum ISP, the level of intrinsic polarization for SN~2005af is $\sim$0.5\% as shown in Fig.~\ref{polint}. The individual values at each epoch are indicated by {\it P}$_{\rm int}$ and $ \theta $$_{\rm int}$ in Table~\ref{log}. The maximum variations of {\it P}$_{\rm int}$ and $ \theta $$_{\rm int}$ are 2.8 and 4.9 times larger than the mean errors of the individual measurements, respectively. Using the Kesteven et al. (\cite{ke76}) criteria,  {\it P}$_{\rm int}$ is non-variable ({\it p} = 27.84\%) and $ \theta $$_{\rm int}$ is possibly variable ({\it p} = 0.65\%). Therefore, the marginal enhanced observed polarization in the plateau phase with a decrease in the nebular phase including the PA rotation is also present in intrinsic polarization. 

Our single {\it V} band measurement on Mar. 12 shows a higher intrinsic polarization level compared to the {\it R} band level on the same night. This may be explained by the effect of addition of (unpolarized) H$\alpha$ line emission which falls in the {\it R} band. This would be consistent with the fact that the SN indeed has intrinsic polarization.  The {\it V}$-${\it R} polarization difference would then represent the intrinsic polarization PA plane, as already indicated (Fig.~\ref{qu}).

\section{Discussion}
\label{comm}

The rotation of the intrinsic PA throughout the plateau phase, as observed in Figs.~\ref{qu}~and~\ref{polint}, is interesting and requires explanation. If the SN's evolving, electron scattering shell had a constant symmetry, the ensuing polarization would vary according to the envelope's size and optical depth, for a given axial ratio (assuming, for simplicity, that such an envelope has an ellipsoidal shape; see below). However in this case the polarization changes would trace a line of constant inclination in the {\it Q}$-${\it U} diagram. The fact that the polarization PA changes with time is consistent with scattering regions of enhanced density that evolve with time (Figs.~\ref{evol}~and~\ref{qu}). The ensuing net, observed polarization percentage and PA would then vary with time (Magalh\~aes et al. \cite{ma06}). Simulations have shown that, in order for such optically thick {\it blobs} to have a significant impact on the polarization value, they would have to be relatively large (a sizable fraction of the envelope) and close to or immersed in the envelope (Rodrigues \& Magalh\~aes \cite{ro00}). A plausible origin for such regions would be turbulence likely to exist in the outer SN envelope. Another mechanism is the light echo model (Wang \& Wheeler \cite{wawh96}; Wang et al. \cite{wa96}), which includes polarization by circumstellar dust-scattering {\it blobs}. SN 2004dj also showed PA rotation but within the nebular phase (Leonard et al. \cite{le06}), after the increase in polarization that took place earlier in that phase; the physical reason may hence not be the same but it is another indication that a constant symmetry is probably a simpliflication in the SN environment.

Regarding SN~2005af's intrinsic polarization, we can compare our results with the models of asphericity effects in electron scattering-dominated photospheres of H\"oflich (\cite{ho91}).  SN~2005af's mean level of $\sim$0.5\% and the slight increment in the plateau phase is consistent with an axial ratio {\it E} $\sim$1.2 for the elliptical density distribution of the envelope with an inclination {\it i} $\sim$30$\degr$, assuming that the emission corresponds to a surface of constant brightness (case {\it CS}) as shown in Fig.~8 of H\"oflich (\cite{ho91}). Another model that can be fitted to our results (but with a lower probability) is ({\it E}=0.8, {\it i}=56$\degr$, {\it CS}). According to this, the overall asphericity inferred for SN~2005af is consistent with $\sim$20\%.

The level of intrinsic polarization for SN~2005af seems to be consistent with the fact that no SN II-P event has been found to show an intrinsic polarization greater than 0.5\%$-$0.6\% (see Fig.~2 from Leonard \& Filippenko \cite{le05} and Leonard et al. \cite{le06}). Nevertheless, the temporal evolution for SN~2005af, with a slight enhancement of intrinsic polarization in the plateau phase and a subsequent decline in the early nebular phase, contrasts with other SNe II-P. SN~1999em (Leonard et al. \cite{le01}), for instance, shows a significant increase between the early plateau phase (0.1\%) and the early nebular phase (0.5\%). SN~2004dj shows a null polarization toward the end of the plateau phase and an abrupt appearance of significant polarization ($\sim$0.6\%) when the inner core is revealed $\sim$90~days after the explosion (Leonard et al. \cite{le06}), early in the nebular phase. In the present case, the non-detection (in the {\it R} band) of the expected polarization enhancement in the last observed epoch  ($\sim$60~days after the event, i.e., at the onset of the nebular phase), may be understood if this epoch was not yet within the phase of radioactive decay. Unfortunately, our $\sim$30-day coverage is not long enough to confirm this issue.

Our study shows that time evolution of SN polarization, in particular SNe II-P events, has strong potential to help understand the geometry of these explosions. Efforts to add polarimetric facilities to obtain a longer temporal coverage for these events would be valuable. We believe that a medium-sized, 2m-class robotic telescope (Magalh\~aes et al. \cite{ma05}) with imaging- and spectro-polarimetric capabilities would fit these requirements well.

\begin{acknowledgements}
We thank the anonymous referee for pointing out the paper by Leonard et al. (\cite{le06}) to us, which appeared after we submitted this paper, and his/her useful comments. A. P. is thankful to FAPESP (grant 02/12880$-$0) and A.M.M. to FAPESP and CNPq for financial support. Polarimetry at IAG-USP is supported by FAPESP grant 01/12589$-$1. This research has made use of the VizieR catalogue access tools operated at CDS, Strasbourg, France.

\end{acknowledgements}


\end{document}